\pdfoutput=1
\documentclass{llncs}
\usepackage{makeidx}  

\usepackage{cite}
\usepackage[pdftex]{graphicx}
\usepackage[cmex10]{amsmath}
\usepackage{array}
\usepackage{mdwmath}
\usepackage{mdwtab}
\usepackage{eqparbox}
\usepackage[caption=false,font=footnotesize]{subfig}
\usepackage{fixltx2e}
\usepackage{url}
\usepackage{amssymb}
\usepackage{filecontents}
\usepackage{comment}
\usepackage{url}
\usepackage{balance}
\usepackage{booktabs}
\usepackage[misc]{ifsym}
\usepackage{algorithm}
\usepackage{algorithmic}
\usepackage{multicol}

\begin{document}

\title{Efficient Sequential and Parallel Algorithms for Estimating Higher Order Spectra}
\titlerunning{Efficient Algorithms for HOS}  
%
\author{Abdullah-Al Mamun\inst{1} \and
Zigeng Wang\inst{1} \and
	Xingyu Cai\inst{1} \and
	Nalini Ravishanker\inst{2} \and
	Sanguthevar Rajasekaran\inst{1}$^{(\textrm{\Letter})}$}
\authorrunning{AA Mamun et al.} 
%
\tocauthor{Abdullah-Al Mamun, Zigeng Wang, Xingyu Cai, Nalini Ravishanker, and Sanguthevar Rajasekaran}
\institute{Dept. of Computer Science,
	University of Connecticut, Storrs, CT 06269, USA
\email{rajasek@engr.uconn.edu}
\and
Dept. of Statistics,
University of Connecticut, Storrs, CT 06269, USA}

\maketitle              

\vspace*{-0.3in}
\begin{abstract}
Polyspectral estimation is a problem of great importance in the analysis of nonlinear time series that has applications in biomedical signal processing, communications, geophysics, image, radar, sonar and speech processing, etc. Higher order spectra (HOS) have been used in unsupervised and supervised clustering in big data scenarios, in testing for Gaussianity, to suppress Gaussian noise, to characterize nonlinearities in time series data, and so on \cite{NiPe93}.

\hspace*{0.2in} Any algorithm for computing the $k$th order spectra of a 
time series of length $n$ needs $\Omega(n^{k-1})$ time since the output size will be $\Omega(n^{k-1})$ as well. Given that we live in an era of big data, $n$ could be very large. In this case, sequential algorithms might take unacceptable amounts of time. Thus it is essential to develop parallel algorithms. There is also room for improving existing sequential algorithms. In addition, parallel algorithms in the literature are nongeneric. In this paper we offer generic sequential algorithms for computing higher order spectra that are asymptotically faster than any published algorithm for HOS. Further, we offer memory efficient algorithms. We also present optimal parallel implementations of these algorithms on parallel computing models such as the PRAM and the mesh. We provide experimental results on our sequential and parallel algorithms. Our parallel implementation achieves very good speedups.

\end{abstract}
\vspace*{-0.4in}
\section{Introduction}
\label{sec:intro}
\vspace*{-0.1in}
Fast computation of HOS such as the bispectrum and the trispectrum becomes escpecially crucial for long nonlinear time series. For example, intra-day financial data analysis usually involves very long time series of stock or index returns or time durations between events of interest such as price or volume changes, see \cite{zou2015}.
Typically, depending on the liquidity of a stock, the time series length within a single day can be as high as 20,000 or more. 
\cite{sinha2007} discusses the use of HOS for monitoring the condition of rotating machinery due to cracks whose signatures are captured as long nonlinear time series consiting of 2,560 observations per second. Existing algorithms are very slow. For instance, the MATLAB code \cite{swami2001} to compute the bispectrum takes 23 seconds on an input of size 2,048.
In the application of cracks and misalignment detection \cite{sinha2007}, if we collect samples for one hour, the sample size will be more than 9 million and the MATLAB code will take an estimated time of more than 14 years! Thus it is essential to 
improve existing sequential algorithms, see \cite{Hinich82}. It is also important to develop effective parallel algorithms.  Existing parallel algorithms are either inefficient or apply to only specific architectures. In this paper we offer general parallel algorithms that are very efficient.

\vspace{0.1in}

\noindent{\bf Problem Statement:} If $X(i)$ is a stationary random process ($i$ denoting discrete time), the moments of order $k$ are given by \cite{NiPe93}:
$$
m^X_k(w_1,w_2,\ldots,w_{k-1})=E[X(i)X(i+w_1)X(i+w_2)\cdots X(i+w_{k-1})].$$
Cumulants are functions of the moments. For example, the first order cumulant is defined as
$c_1^X=m^X_1=E[X(i)]$, the second order cumulant as 
$c_2^X(w_1)=m_2^X(w_1)-(m_1^X)^2$, and so on.
The moments and cumulants defined above are based on expectations over the (infinite) ensemble.
For ergodic processes, these ensemble averages may be estimated using the corresponding time averages.
The Fourier transform of the third and fourth cumulants are respectively the bispectrum and the trispectrum. 
The problem we address is the following: Given a finite sequence $X(1),X(2),\ldots,X(n)$, compute 
smoothed sample bispectrum and trispectrum which are  statistically consistent estimates of the corresponding true HOS. 

\vspace*{0.1in}
\noindent{\bf Some Applications:} HOS are useful in unsupervised and supervised classification of long sequences of nonlinear time series with applications in finance, geoscience, neuroscience, etc. HOS can also be used as a test for Gaussianity of any data, since if $X(i)$ is Gaussian, the cumulant $c_k^X(w_1,w_2,\ldots,w_{k-1})=0$ for $k>2$ \cite{NiPe93,sraogabr}. HOS can also be used to test for linearity of any data. Other applications include characterization of coronary artery disease \cite{achar2017}, analysis of breast thermograms \cite{ANS+14}, communication systems \cite{san2013}, etc. 

In this paper, a major part is devoted to a discussion on computing the bispectrum. 
However the techniques proposed extend to 
trispectra as well (as we explain toward the end).

\vspace*{0.05in}
\noindent{\bf Direct Method for HOS:}
Two kinds of algorithms can be found in the literature for HOS: direct method using the Fast Fourier Transform (FFT) and an indirect method via the Fourier transform of the third cumulant. In this paper we use the direct method. However, the algorithms we propose can be used for the indirect method as well. The following summary of the direct method can be found in \cite{NiPe93}.
Let $X(1),X(2),\ldots,X(n)$ be the input sequence. The direct bispectrum (DBS) method works as follows:\\

\vspace*{-0.3in}
\begin{center}
	{\bf Algorithm} DBS
\end{center}
\vspace*{-0.1in}
\hspace*{0.2in} 1) Partition the input into $K$ parts with $M$ samples in each part. Let $X_i$ stand for the $i$th part, for $1\leq i\leq \frac{n}{M}$.\\
\hspace*{0.2in} 2) In each part subtract the mean of that part from each element in the part.\\
\hspace*{0.2in} 3) Compute the Discrete Fourier Transform $F_X^i(k)$ for each part:
$F_X^i(k)=\sum_{u=0}^{M-1}X^i(u)e^{-j\frac{2\pi}{M}uk},$
${\rm for}\ k=0,1,\ldots,M-1;i=1,2,\ldots,K$, and $j=\sqrt{-1}$.\\
\hspace*{0.2in} 4) Estimate the  
raw bispectrum of each part as:\protect\linebreak
$C_3^{X_i}(k_1,k_2)=\frac{1}{M}F_X^i(k_1)F_X^i(k_2)F_X^i(k_1+k_2),$
${\rm for}\ i=1,2,\ldots,K.$
Due to various symmetries, it suffices to compute the bispectrum $C_3^{X_i}(k_1,k_2)$ only in the principal domain: 
$0\leq k_2\leq k_1,k_1+k_2<M/2$.\\
\hspace*{0.2in} 5) This step performs some smoothing over a window of size $M_3\times M_3$ and yields a consistent estimate of the true bispectrum: \\
$\widetilde{C}_3^{X_i}(k_1,k_2)=\frac{1}{M_3^2}\sum_{n_1=-M_3/2}^{M_3/2-1}\sum_{n_2=-M_3/2}^{M_3/2-1}C_3^{X_i}(k_1+n_1,k_2+n_2).$\\
\hspace*{0.2in} 6) The estimated bispectrum of the entire time series is computed as the average over all parts:
$\hat{C}(w_1,w_2)=\frac{1}{K}\sum_{i=1}^K\tilde{C}^{X_i}_3(w_1,w_2)$. 

\vspace*{0.05in}

\noindent{\bf Time Complexity Analysis:} Step 2 in the direct method takes $O(n)$ time. Step 3 takes a total of $O(n\log M)$ time. Step 4 takes $O(M^2)$ time per part. Thus Step 4 takes a total of $O(KM^2)=O(Mn)$. In Step 5 smoothing is done. For every point $(k_1,k_2)$, the smoothed value $\widetilde{C}_3^{X_i}(k_1,k_2)$ is computed as the average value of $C_3^{X_i}$ over a region of size $O(M_3^2)$. Thus each such computation takes $O(M_3^2)$ time. The total time taken in Step 5 is $O(M^2KM_3^2)=O(MnM_3^2)$. 

In summary, the total run time of the direct method is $O(MnM_3^2)$. In this paper we show that this run time can be improved to $O(Mn)$. Note that this run time is independent of $M_3$.

\vspace{0.05in}
\noindent{\bf Known parallel algorithms for HOS:} We summarize below some of the the known algorithms. As we can see, these algorithms are very inefficient and restricted to specific architectures.

Manolakos, et al. \cite{MSB91} discuss the importance of power spectra in signal processing. Followed by this, they employ the canonical mapping methodology (CMM) to derive parallel programs for computing bispectrum. This paper focused exclusively on the design of the systolic array and no experimental results were presented.
In \cite{KaMa95}, the authors present data parallel algorithms for computing 3rd and 4th order moments on the MasPar-1 SIMD parallel system. Their program handles input sequences of length up to $2^{10}$.  Their algorithm can be thought of as a mesh algorithm. No time complexity analyses were given in the paper and the algorithm was very specific for the MasPar-1 machine. 
In \cite{LED99} also, the authors consider the parallel computation of bispectrum. They have implemented the direct and the indirect methods using two different parallel programming techniques: semi-automatic and fully automatic using the Power C Analyzer. The machine used was the Silicon Graphics Power Challenge MIMD Machine HOTBLACK. This paper also falls under the category of developing a parallel program for a specific machine. In \cite{hajmh2015} and \cite{hajmh2016} the authors consider parallel reconstruction of images using bispectra. They parallelize the bispectrum algorithm in a straight forward manner without worrying about achieving optimal run times.

\vspace{0.05in}

\noindent{\bf Contributions of this paper:}
None of the above papers deals with the problem of constructing smoothed sample HOS which are consistent estimates of the true HOS. In this paper our focus is on developing generic parallel algorithms that can be employed on any parallel machine or platform. We also provide experimental evaluations of our algorithms. For HOS computing algorithms one of the major bottlenecks could be in the memory needed. For computing order $k$ moments the memory needed is $\Omega(n^{k-1})$. This could indeed be prohibitive. For example, when $k=3$ and $n=10^6$, the memory needed will be at least 1,000 GB. Thus it is essential to develop memory efficient algorithms. In this paper we address this crucial problem. Also, for bispectrum computation with smoothing over a window of size $M_3$, existing algorithms take $O(nMM_3^2)$ time. In this paper we present sequential and parallel algorithms that do only $O(nM)$ work. Here $M$ is the size of each part of the input. 

\vspace*{-0.05in}
\section{A Better Algorithm for the Direct Method}
\vspace*{-0.15in}
In this section we show how to improve the run time of the direct method from $O(MnM_3^2)$ to $O(Mn)$.
The new algorithm is based on an efficient way of computing window sums that we describe next.

\vspace*{-0.2in}
\subsection{Computing window sums}
\vspace*{-0.1in}
{\bf The case of 1D data:} Let $X=k_1,k_2,\ldots,k_n$ be any sequence of real numbers and let $w$ be a window size. The problem is to compute $s_i=\sum_{j=1}^w k_{i+j-1}$, for $1\leq i\leq (n-w+1)$. 

A straight forward algorithm for this problem will take $O(nw)$ time. We can improve this to $O(n)$ using overlaps in successive window sums. Specifically, $s_{i+1}=s_i-k_i+k_{i+j}$, for $1\leq i\leq (n-w)$. This means that $s_{i+1}$ can be obtained from $s_i$ in $O(1)$ time. Therefore, if we compute the window sums in this order: $s_1,s_2,\ldots, s_{n-w+1}$, then we can compute all of them in $O(n)$ time.

\vspace*{0.1in}

\noindent{\bf The case of 2D data:} The above idea can be extended to 2D data as well. Let $A=(a_{i,j})$ be an $n\times n$ matrix and let $w$ be a window size.  Consider the problem of computing $s_{i,j}=\sum_{u=1}^w\sum_{v=1}^wa_{i+u-1,j+v-1}$, for $1\leq i\leq (n-w+1)$ and $1\leq j\leq(n-w+1)$.

A trivial algorithm for solving the above problem will take $O(n^2w^2)$ time. We can improve this run time to $O(n^2)$ as follows.

\vspace*{-0.1in}
\begin{algorithm}[h]
\caption{WS}
\label{alg-ws}
\begin{multicols}{2}
\begin{algorithmic}[1]
\FOR{$i=1$ {\bf to} $(n-w+1)$}
  \FOR{$j=1$ {\bf to} $(n-w+1)$}
    \STATE  $r_{i,j}=\sum_{k=0}^{w-1} A_{i,j+k}$;
  \ENDFOR
\ENDFOR
\FOR{$i=1$ {\bf to} $(n-w+1)$}
  \FOR{$j=1$ {\bf to} $(n-w+1)$}
    \STATE  $c_{i,j}=\sum_{k=0}^{w-1} A_{i+k,j};$
  \ENDFOR
\ENDFOR
\STATE Compute $s_{1,1}$ in $w^2$ time;
\FOR{$j=2$ {\bf to} $(n-w+1)$}
  \STATE  $s_{1,j}=s_{1,j-1}-c_{1,j-1}+c_{1,j+w-1}$;
\ENDFOR
\FOR{$i=2$ {\bf to} $(n-w+1)$}
  \FOR{$j=1$ {\bf to} $(n-w+1)$}
    \STATE   $s_{i,j}=s_{i-1,j}-r_{i-1,j}+r_{i+w-1,j}$;
  \ENDFOR
\ENDFOR
\end{algorithmic}
\end{multicols}
\end{algorithm}

\noindent{\bf Analysis:} The total run time of the above algorithm is $O(n^2)$. 

\vspace*{-0.1in}
\subsection{Direct Method for Bispectrum}
\vspace*{-0.1in}
We can employ the above window sums algorithms in the smoothing step (5) of the direct method. In this case we get the following theorem.
\begin{theorem}\label{directseq}
	We can compute bispectrum of any input of size $n$ using the direct method in $O(Mn)$ time, $M$ being the partition size. $\Box$
\end{theorem}

\vspace*{-0.2in}
\section{Parallel Models and Preliminaries}
\label{sec:other}
\vspace*{-0.1in}
In this section we describe the parallel models of computing that we employ in this paper, namely, the PRAM and the mesh. A Parallel Random Access Machine (PRAM) is a collection of RAMs working in synchrony where communication takes place with the help of a common block of shared memory \cite{Jaj92,HSR98}.  Depending on how read and write conflicts are handled, a PRAM can further be classified into three: Exclusive Read and Exclusive Write (EREW) PRAM, Concurrent Read and Exclusive Write (CREW) PRAM, and Concurrent Read and Concurrent Write (CRCW) PRAM.
There are variants of a CRCW PRAM depending on how write conflicts are handled. In a Common-CRCW PRAM, concurrent writes are permissible only if the processors trying to write in the same cell at the same time have the same data to write.  In an Arbitrary-CRCW PRAM, if more than one processor tries to write in the same cell at the same time, an arbitrary one of them succeeds.  In a Priority-CRCW PRAM, processors have assigned priorities. Write conflicts are resolved using these priorities.

An $n \times n$ mesh can be represented as a directed $n\times n$ grid-graph whose nodes correspond to processing elements and whose edges correspond to bidirectional communication links \cite{HSR98}.  If two processors are connected by an edge, they can communicate in a unit step. Otherwise, they communicate by sending a message along a connecting path. The work done by a parallel algorithm that uses $P$ processors and runs in time $T$ is defined as the product $P\times T$.

Let $\oplus$ be any associative unit-time computable binary operator defined in some domain $\Sigma$. Given a sequence of $n$ elements $k_1,k_2,\ldots,k_n$ from $\Sigma$, the problem of {\em prefix computation} is to compute $k_1,k_1\oplus k_2,
\ldots,k_1\oplus k_2\oplus\cdots \oplus k_n$. Proof of the following Lemma can be found in relevant texts (such as \cite{Jaj92,HSR98}).

\vspace*{-0.1in}
\begin{lemma}
	\label{prefix1}
	Prefix computation on a sequence of $n$ elements can be performed in
	$O(\log n)$ time using $\frac{n}{\log n}$ EREW PRAM processors.
\end{lemma}

\vspace*{-0.25in}
\subsection{Window sums on the PRAM}
\vspace*{-0.1in}
We now show how to implement the direct method on an EREW PRAM optimally. First we consider the computation of window sums in 1D and 2D. 

\vspace*{0.1in} 

{\bf The case of 1D data in parallel:} Let $X=k_1,k_2,\ldots,k_n$ be any sequence of real numbers and let $w$ be a window size. The problem is to compute $s_i=\sum_{j=1}^w k_{i+j-1}$, for $1\leq i\leq (n-w+1)$. 

A straight forward PRAM algorithm for this problem could use $(n-w+1)$ processors. Each processor can in parallel compute one window sum in $O(w)$ time. The work done will be $O(nw)$. We can improve these bounds using the prefix computation.
\vspace*{0.1in}

\noindent\hspace*{0.1in} 1) Perform a prefix sums computation on $k_1,k_2,\ldots,k_n$.\\
\hspace*{0.6in} Let the results be $q_1,q_2,\ldots,q_n$; Let $q_0=0$;\\
\hspace*{0.1in} 2) {\bf for} $i=1$ {\bf to} $(n-w+1)$ {\bf in parallel do}\\
\hspace*{0.1in} 3) \hspace*{0.2in} $s_i=q_{i+w-1}-q_{i-1};$

\vspace*{0.1in}
\noindent{\bf Analysis:} Step 1 can be done using $\frac{n}{\log n}$ EREW PRAM processors in $O(\log n)$ time (c.f. Lemma~\ref{prefix1}). The {\bf for} loop of line 2 can be performed in $O(1)$ time using $n$ EREW PRAM processors. Using the slow-down lemma (see e.g., \cite{Jaj92,HSR98}), Step 2 can also be completed in $O(\log n)$ time using $\frac{n}{\log n}$ processors. Thus we arrive at the following lemma.

\vspace*{-0.1in}
\begin{lemma}\label{1Dws}
	The window sums computation problem on any input sequence of length $n$ can be solved in $O(\log n)$ time using $\frac{n}{\log n}$ EREW PRAM processors. $\Box$
\end{lemma}

\vspace*{-0.1in}
\noindent{\bf The case of 2D data in parallel:}
Let $A=(a_{i,j})$ be an $n\times n$ matrix and let $w$ be a window size. We are interested in computing $s_{i,j}=\sum_{u=1}^w\sum_{v=1}^wa_{i+u-1,j+v-1}$, for $1\leq i\leq (n-w+1)$ and $1\leq j\leq(n-w+1)$.

A trivial algorithm for solving the above problem will do $O(n^2w^2)$ work. We can improve this work to $O(n^2)$ as follows. In this algorithm, $t_{i,0}=0$, for $1\leq i\leq(n-w+1)$.

\begin{algorithm}[h]
\caption{WS\_PRAM}
\label{alg-wspram}
\begin{multicols}{2}
\begin{algorithmic}[1]
\FOR{$j=1$ {\bf to} $n$ {\bf in parallel}} 
  \STATE  Compute window sums in column $j$;
  \STATE Specifically, let $c_{i,j}=\sum_{k=0}^{w-1} a_{i+k,j}$, for $1\leq i\leq (n-w+1)$;
\ENDFOR
\FOR{$i=1$ {\bf to} $(n-w+1)$ {\bf in parallel}}
  \STATE Perform a prefix sums computation on $c_{i,1},c_{i,2},\ldots,c_{i,n}$;
  \STATE Let the results be $t_{i,1},t_{i,2},\ldots,t_{i,n}$;
\ENDFOR
\FOR{$i=2$ {\bf to} $(n-w+1)$ {\bf in parallel}}
  \FOR{$j=1$ {\bf to} $(n-w+1)$ {\bf in parallel}}
    \STATE  $s_{i,j}=t_{i,j+w-1}-t_{i,j-1}$;
  \ENDFOR
\ENDFOR
\end{algorithmic}
\end{multicols}
\end{algorithm}

\noindent{\bf Analysis:} In line 1, for a specific value of $j$, window sums can be computed in $O(\log n)$ time using $\frac{n}{\log n}$ EREW PRAM processors (c.f. Lemma~\ref{1Dws}). Thus the {\bf for} loop of line 1 can be completed in $O(\log n)$ time given $\frac{n^2}{\log n}$ EREW PRAM processors. 

In line 5, for any given value of $i$, prefix sums computation can be performed in $O(\log n)$ time using $\frac{n}{\log n}$ EREW PRAM processors (c.f. Lemma~\ref{prefix1}). As a result, the {\bf for} loop of line 5 takes $O(\log n)$ time given $\frac{n^2}{\log n}$ EREW PRAM processors.

Line 11 can performed (for a given $i$ and $j$) in $O(1)$ time using one processor. Therefore, the {\bf for} loop of line 9 can be performed in $O(1)$ time given $(n-w+1)^2$ EREW PRAM processors. Using the slow-down lemma,  the {\bf for} loop of line 9 can also be completed in $O(\log n)$ time using $\frac{n^2}{\log n}$ processors.

Put together, the above algorithm runs in a total of $O(\log n)$ time using $\frac{n^2}{\log n}$ EREW PRAM processors. Clearly, this algorithm is asymptotically work-optimal. We arrive at the following lemma:

\begin{theorem}\label{2Dws}
	The window sums computation problem can be solved in $O(\log n)$ time using $\frac{n^2}{\log n}$ EREW PRAM processors. $\Box$
\end{theorem}

\vspace*{-0.2in}
\subsection{Direct method for bispectrum on a PRAM} 
In this section we present a PRAM algorithm for direct bispectrum computation. There are 5 steps in the algorithm (c.f. Algorithm DBS). We discuss how to parallelize each step.
Let $X(1),X(2),\ldots,X(n)$ be the input sequence. 

Step 1 is that of partitioning the data into $K$ parts and this does not cost any time since the input will be given in the common memory. Let the parts be $X^i,\ 1\leq i \leq K$. 

In Step 2, finding the mean of $X_i$ can be done in $O(\log M)$ time using $\frac{M}{\log M}$ processors, for a specific $i$. Thus the mean of all the parts can be found in $O(\log M)$ time using $\frac{n}{\log M}$ processors. Using the slow down lemma, Step 2 can be performed in $O(\log n)$ time using $\frac{n}{\log n}$ EREW PRAM processors.

Step 3 involves the computation of the Discrete Fourier Transform (DFT) $F_X^i(k)$ for each part:
$F_X^i(k)=\sum_{u=0}^{M-1}X^i(u)e^{-j\frac{2\pi}{M}uk},$ for $ k=0,1,\ldots,M-1;i=1,2,\ldots,K.$ For each part, the time taken is $O(\log M)$ using $M$ processors (see e.g., \cite{Jaj92,HSR98}). Therefore, the DFT of all the parts can be computed in $O(\log M)$ time using $n$ processors. 

We have to estimate the third order spectrum of each part in Step 4. Specifically, we have to compute $C_3^{X_i}(k_1,k_2)$, for $1\leq i\leq K$ and $0\leq k_2\leq k_1,k_1+k_2<M/2$.
This can be done in $O(1)$ time using $O(nM)$ processors. Equivalently, Step 4 can also be done in $O(\log n)$ time using $\frac{nM}{\log n}$ EREW PRAM processors (using the slow down lemma).

Step 5 is concerned with the smoothing operation. The value of the bispectrum at any point is computed as an average over a surrounding window of size $M_3\times M_3$. This Step can be performed using the Algorithm WS\_PRAM (c.f. Theorem~\ref{2Dws}). For each part, this Step can be completed in $O(\log M)$ time using $\frac{M^2}{\log M}$ processors. For all the $K$ parts together, Step 5 takes $O(\log M)$ time using $\frac{nM}{\log M}$ processors. Using the slow down lemma, Step 5 can be completed in $O(\log n)$ time employing $\frac{nM}{\log n}$ processors.

In Step 6, the bispectrum is computed as the average over all parts. In particular, we have to compute 
$\hat{C}(w_1,w_2)=\frac{1}{K}\sum_{i=1}^K\tilde{C}^{X_i}_3(w_1,w_2)$.
 For a given $w_1$ and $w_2$, $\hat{C}(w_1,w_2)$ can be computed using a prefix sums computation on $K$ elements and hence can be done in $O(\log K)$ time using $\frac{K}{\log K}$ processors. Thus Step 6 can be completed in $O(\log K)$ time using $\frac{M^2K}{\log K}=\frac{nM}{\log K}$ processors.  The slow down lemma implies that Step 6 can also done in $O(\log n)$ time using $\frac{nM}{\log n}$ processors.

In summary, we get the following theorem.

\begin{theorem}\label{directpram}
	We can compute the bispectrum on any sequence of length $n$ in $O(\log n)$ time using $\frac{nM}{\log n}$ EREW PRAM processors, where $M$ is the size used to partition the input sequence. $\Box$
\end{theorem}

The following theorems pertain to computing the bispectrum computation in a memory efficient manner. Proofs are omitted due to space constraints and will be supplied in the full version.
\begin{theorem}\label{wsum1}
	We can solve the window sums problem on any $n\times n$ matrix in $O(n\log n)$ time using $\frac{n}{\log n}$ EREW PRAM processors using only $O(nw)$ memory, $w$ being the window size. $\Box$
\end{theorem}

\begin{theorem}\label{mm3}
	Bispectrum computation on any given sequence of length $n$ can be computed in $O(n\log n)$ time using $\frac{M}{\log n}$ EREW PRAM processors and $O(MM_3)$ memory, where $M$ is the size of each part and $M_3$ is the window size of smoothing (assuming that $M=\omega(\log n)$). $\Box$
\end{theorem}

\begin{theorem}\label{wsum2}
	Window sums on any $n\times n$ data can be computed in $O\left(\frac{n^2}{w^2}\log n\right)$ time using $\frac{w^2}{\log n}$ EREW PRAM processors and $O(w^2)$ memory, $w$ being the window size. $\Box$
\end{theorem}

\begin{theorem}\label{wsum3}
	Window sums on any $n\times n$ data can be computed in $O(n^2\log n)$ time using $\frac{w}{\log n}$ EREW PRAM processors and $O(w)$ memory, $w$ being the window size. $\Box$
\end{theorem}

Note that the work done in the above algorithm is $O(n^2w)$ and hence the algorithm is not work optimal. However, the memory used is very small. Theorems \ref{2Dws}, \ref{wsum1}, \ref{wsum2}, and \ref{wsum3} consider memories of different specific sizes. Theorems \ref{2Dws} and \ref{wsum2} can be used to develop work optimal algorithms when the memory available is $m$ for any $w^2\leq m\leq n^2$. The following theorems consider the mesh model and higher order spectra, respectively. Proofs are omitted due to space constraints.

\begin{theorem}
	Bispectrum computation of a sequence of length $n$ can be performed on an $n\times n$ mesh in $O(n)$ time. $\Box$
\end{theorem}

\begin{theorem}
	On any input of size $n$, we can compute $k$th order spectrum in $O(nM^{k-2})$ time, for any $k\geq 3$ where $M$ is the size of each part in the input. $\Box$
\end{theorem}

\vspace*{-0.2in}
\section{Experimental Results}
\vspace*{-0.15in}
We have conducted extensive experiments to evaluate the performance of our proposed approaches. In this section we report the results.

\vspace*{-0.2in}
\subsection{Test Platform}
\vspace*{-0.1in}
All the experiments have been performed on the test server, which is equipped with Intel(R) Xeon(R) CPU E5-2667 v3 @ 3.20GHz, with 16 cores (Hyperthreading to 32 threads), 256 GB main memory and 4 TB HDD disk.
All the algorithms have been implemented using C++ and the standard GCC compiler. The parallel version is implemented using OpenMP.
We have used a value of $K=1$ throughout. We have generated different types of the time series data sequences for our experiments using guidelines given in \cite{NaliniBispec, jones1978}. 

We have implemented algorithms for spectral computation for bispectrum and trispectrum computations. 
For both of them, we have compared 5 different approaches: Naive approach with $O(n^2m^2)$ run time, denoted as \textbf{Naive}. Here $m$ is nothing but $M_3$; 
Our sequential algorithm that takes $O(n^2)$ time (c.f. Theorem~\ref{directseq}) - Call this algorithm as \textbf{WS} in consistent with above sections; 
Our fastest algorithm of Theorem~\ref{mm3} that does $O(n^2)$ work and uses $O(nm)$ space - Call this algorithm \textbf{Fast}; 
The most efficient algorithm in both time and memory ($O(n^2)$ time, $O(m^2)$ memory) - Call this algorithm \textbf{Efficient}; Parallel approach (denoted as \textbf{Parallel}) with $P$ threads, $P=2,4,8,16$.

\vspace*{-0.1in}
\subsection{Run Time and Memory Comparisons}
\begin{figure}[!ht]
	\centering
	\includegraphics[width = 2.3 in]{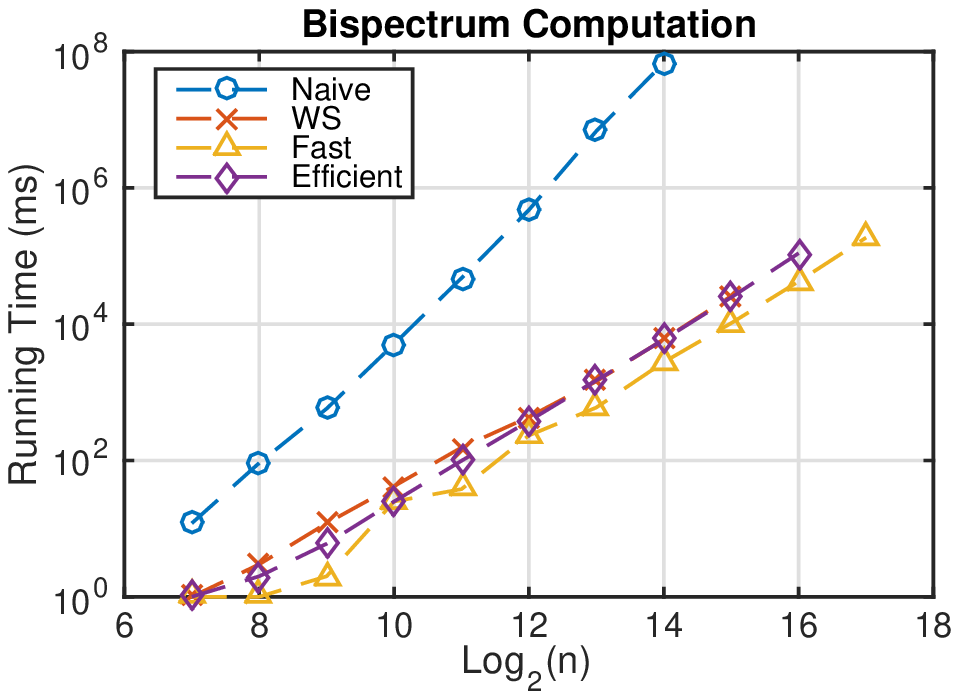}
	\includegraphics[width = 2.3 in]{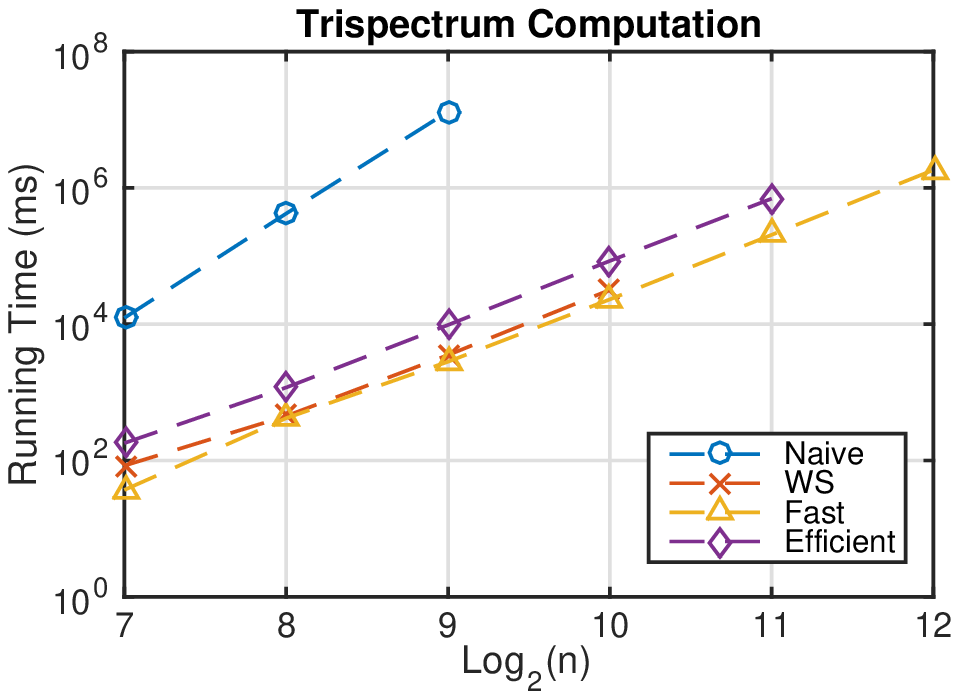}
	\caption{Run time comparison for different values of $n$}
	\label{fig-time}
\end{figure}

\vspace*{-0.12in}
We have set a run time threshold of 10 hours and a memory threshold of 100 GB. Any algorithm exceeding one or both of these thresholds was forced to stop. For large datasets such as those with $n=2^{14}, 2^{15}$, some of the algorithms exceeded these thresholds.
In Figure~\ref{fig-time}, we show the running time of different approaches for bispectrum and trispectrum, respectively. Note that this is a log plot. Thus the parallel curves show orders of magnitude difference. From this figure we can clearly see that compared with the naive algorithm, all of our algorithms offer much better run times. 

\vspace*{-0.2in}

\vspace*{-0.2in}

We provide the maximum memory cost during the running time of each algorithm in Table~\ref{tab-mem}.
From this table we see that the memory cost in our experiments is consistent with our theoretical analyses.
As it is shown, the \textbf{Efficient} approach is extremely memory efficient. 
For instance, for the time series sequence with a length of $n=2^{16}$, it only requires less than $100$ MB of memory.
In contrast, even the second memory efficient approach \textbf{Fast} uses around 1 GB, and the others occupy more than 30 GB.
\textbf{Efficient} would take a longer time than \textbf{Fast}, which demonstrates the trade-off between space and time.


\begin{table}[!ht]
	\centering
	\footnotesize
	\caption{Memory (in MB) Comparison}
	\label{tab-mem}
	\begin{tabular}{|c|c|c|c|c||c|c|c|c|c|} \hline
		$n$     & {\bf Naive}  & {\bf WS}  & {\bf Fast} & {\bf Efficient} & $n$     & {\bf Naive}  & {\bf WS}  & {\bf Fast}      & {\bf Efficient} \\ \hline
		\multicolumn{5}{|c||}{Bispectrum} & \multicolumn{5}{c|}{Trispectrum}                     \\ \hline
		$2^{12}$ & 504.8   & 504.7   & 14.5   & 5.2 & 		$2^7$    & 56.8  & 56.7  & 8.4     & 3.6 \\
		$2^{13}$ & 2030.7  & 2030.5  & 38.6   & 9.0 & 		$2^8$    & 448.0   & 448.0   & 38.2     & 7.3 \\
		$2^{14}$ & 8176.3 & 8176.3 & 113.8  & 17.6 & 		$2^9$    & 3682.7 & 3682.7 & 215.8     & 19.3 \\
		$2^{15}$ & NA      & 32890.0 & 344.3 & 37.9  & 		$2^{10}$ & NA  & 30224.3 & 1297.7  & 64.6 \\
		$2^{16}$ & NA      & NA      & 1055.8 & 86.3  & 		$2^{11}$ & NA    & NA      & 7869.3  &  228.4 \\ \hline
	\end{tabular}
\end{table}


We have compared running times of bispectrum computation by our \textbf{Fast} implementation and HOSA Toolbox in MATLAB \cite{swami2001} for single thread. {\bf The results from the two programs match exactly}. Table~\ref{tab-comparison} shows how fast our \textbf{Fast} implementation becomes when series lengths increase. We have computed bispectrum for every pair of frequencies with linear smoothing window using both of these implementations . In our experiments with HOSA Toolbox, we supplied $0.0$ for overlap value and series length as segment size.

\vspace*{-0.2in}
\begin{table}[!ht]
	\centering
	\footnotesize
	\caption{Comparison of running times (in sec) of {\bf Fast} and {\bf HOSA Toolbox}}
	\label{tab-comparison}
	\begin{tabular}{| c |c | c | c |} \hline
		series length & window length    & {\bf Fast}  & {\bf HOSA Toolbox} \\ \hline
		128 & 21 & 0.001 & 0.010 \\
		256 & 33 & 0.005 & 0.042 \\
		512 & 49 & 0.011 & 0.220 \\
		1,024 & 77 & 0.032 & 1.755 \\
		2,048 & 117 & 0.126 & 23.342 \\
		4,096 & 181 & 0.448 & 329.961 \\
		8,192 & 279 & 1.751 & 3102.4 \\ \hline
	\end{tabular}
\end{table}
\vspace*{-0.45in}

\subsection{Multi-core Parallel Approach Evaluation}
Next, we evaluate our proposed parallel algorithm.
Due to the fact that the \textbf{Efficient} approach has a significant advantage in memory, we have implemented the parallel version of \textbf{Efficient} to offer a fast and memory efficient approach in high order spectra computations.
We have tested the {\bf Parallel} algorithm using $P=2,4,8,16$ and $n$ from $2^7$ to $2^{15}$. 

\vspace*{-0.2in}

\begin{figure}[!ht]
	\centering
	\includegraphics[width = 2.3 in]{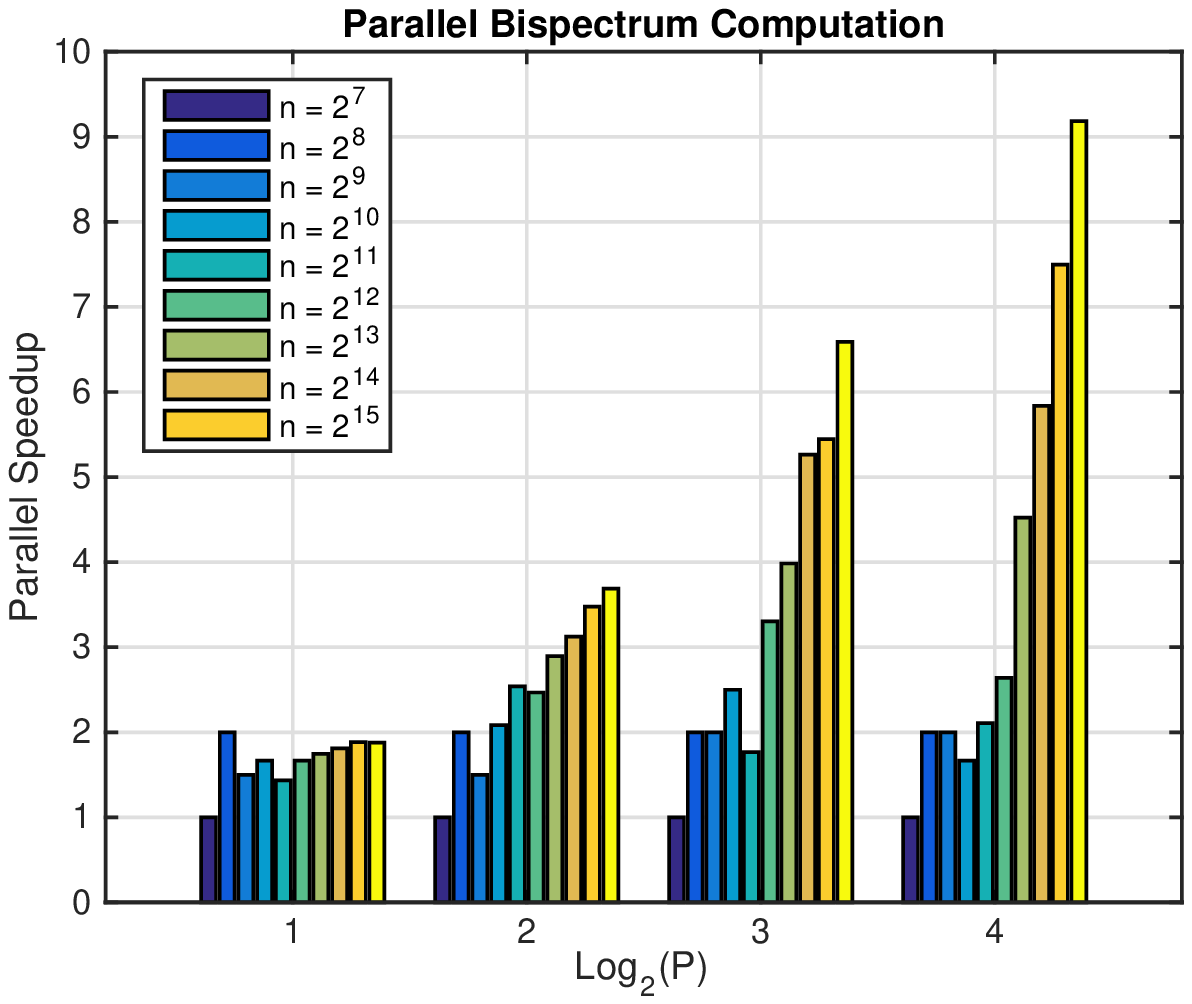}
	\includegraphics[width = 2.3 in]{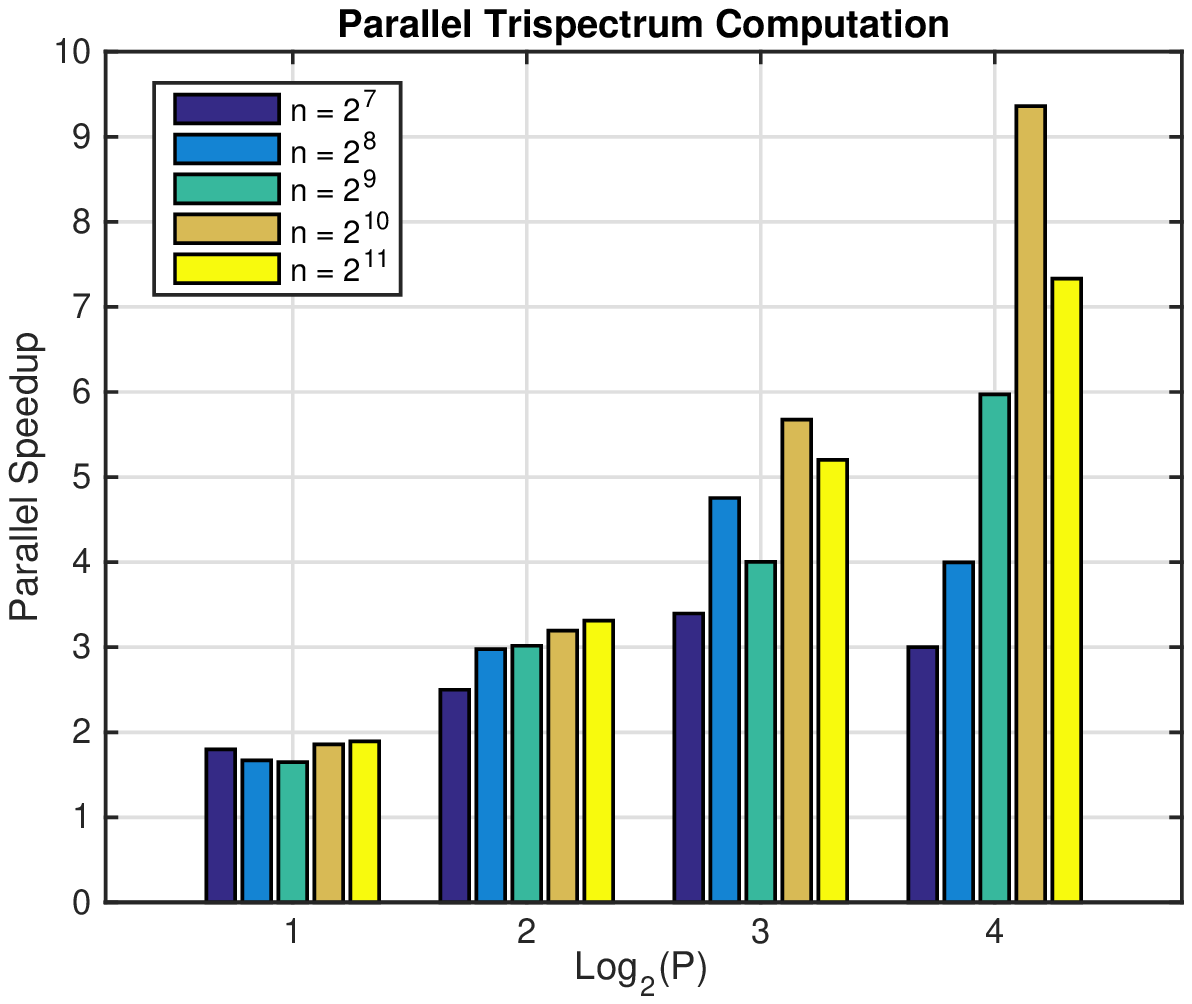}
	\caption{Speedups using 2, 4, 8, 16 cores}
	\label{fig-paraspeedup}
\end{figure}

\vspace*{-0.2in}

In Figure~\ref{fig-paraspeedup} we plot the parallel speedup against number of cores $P$. 
As we can see, from 2 cores to 16 cores ($\log_2(P)=1,2,3,4$), the speedup is increasing if the number of cores is increasing. Sometimes super-linear speedup could also be achieved for 2 cores and 4 cores.

Another interesting fact is that for small data lengths (small $n$ values), the gain of multi-cores is not as significant as for larger $n$ values. This is due to the overhead introduced in multi-core implementations, such as processor scheduling and communication. 
For instance, in the case of $n=2^7$ of bispectrum computation, our 2-core {\bf Efficient} approach has a run time of around 1 ms. However, using 16-threads {\bf Parallel} still took 1 ms to finish.
Thus for small datasets, the overhead of work scheduling becomes dominant.
On the contrary, the computation time is still the dominant part for large datasets, e.g., more than 10 hours for $n=2^{15}$ using the sequential algorithm. This makes our parallel algorithm especially useful for large datasets.

From the memory point of view, the memory cost for the parallel implementation is linearly dependent on the number of cores. This is due to the fact that each processor is independently working on its own smoothing window. 

\vspace*{-0.2in}
\subsection{Summary of Experiments}
\vspace*{-0.1in}
We have evaluated four approaches, {\bf Naive, WS, Fast}, and {\bf Efficient}, respectively, as well as the {\bf Parallel} algorithm. Both bispectrum and trispectrum implementations have been tested on different lengths of time series data. 

All of our proposed algorithms outperform the {\bf Naive} algorithm by orders of magnitude, in terms of both run time and memory. Please note that the naive algorithm is the best found in the literature. {\bf Fast} is $25\times 10^3$ times faster when $n=2^{15}$, and {\bf Efficient} uses less than $1/200$th of the memory used by the naive algorithm. Even though {\bf WS} is simpler, {\bf Fast} runs the fastest. 
It could be due to the cache misses and memory accessing time costs, as {\bf WS} occupies a significantly larger memory. {\bf Fast} and {\bf Efficient} display a memory-time trade-off. 
{\bf Efficient} has a better balance and is extremely frugal in memory usage.

{\bf Parallel} is a fast and memory saving algorithm, suitable for problems with very large $n$. A linear speedup can be achieved by {\bf Parallel} on larger datasets. The memory cost for {\bf Parallel} is also linearly dependent on $P$. For large $n$, {\bf Parallel}'s performance is better than for small $n$. This is due to the overhead of parallel implementation, making the parallel approach more preferred for large datasets. Large data sets are quite relevant in today's world of big data.

\vspace*{-0.2in}
\section{Conclusions}
\vspace*{-0.1in}
In this paper we offer efficient sequential and parallel algorithms for computing HOS. The work done by our algorithms is asymptotically better than any known algorithm for HOS. 
For our proposed sequential approach, the run time is reduced to $O(n^2)$.
Another crucial problem in computing HOS is in the need for large memories.
To address this problem, we offer memory efficient algorithms. We have also presented work optimal parallel algorithms. 
Experimental results reveal that our algorithms are indeed highly competitive and especially suitable for long sequence data problems.

\vspace*{-0.2in}
\section*{Acknowledgement}
\vspace*{-0.1in}
This work has been partly supported by NSF Grants 1447711 \& 1743418 to SR.

\vspace*{-0.2in}



%

\end{document}